\documentclass[12pt]{article}
\usepackage{dsfont}
\usepackage{amssymb}
\usepackage{mathrsfs}

\begin{document}
\title{Towards a classification of wave catastrophes}
\author{
T. Kiss$^{1,2}$ and U. Leonhardt$^{3}$\\
$^{1}$Research Institute for Solid State Physics and Optics,\\ 
H-1525 Budapest, P.~O.~Box 49, Hungary\\ 
$^{2}$Institute of Physics, University of P\'ecs,\\
Ifj\'us\'ag u.\ 6. H-7624 P\'ecs, Hungary\\
$^{3}$School of Physics and Astronomy, University of St Andrews,\\
North Haugh, St Andrews KY16 9SS, Scotland
}
\date{}
\maketitle
\begin{abstract}
Wave catastrophes are characterized by logarithmic phase singularities. Examples are light at the horizon of a black hole, sound in transsonic fluids, waves in accelerated frames, light in singular dielectrics and slow light close to a zero of the group velocity. We show that the wave amplitude grows with a half-integer power for monodirectional and symmetric wave catastrophes.\\[2mm]
\end{abstract}
\newpage

Imagine light propagating away from a black hole. Suppose that the light has been emitted immediately before the horizon. A distant observer decomposes the light into its spectral components with frequencies $\omega$. The wavelength of each spectral component must shrink to zero close to the horizon, in order to compensate for the infinite gravitational redshift here. It turns out \cite{Brout} that the wavelength is proportional to the radial distance from the horizon. Consequently, the wavenumber depends inversely on the distance and the phase diverges logarithmically. Such rapid oscillations occur only in the direction orthogonal to the horizon, which allows us to ignore all other spatial dimensions and to focus on an effectively 1+1 dimensional model where we can also ignore the polarization of light. In this model a monochromatic wave $\varphi$ behaves like
\begin{equation}
\varphi \sim (z-z_0)^{i\nu} e^{-i\omega t}
\label{eq:wave}
\end{equation}
near the horizon at $z_0$ where $z$ denotes the spatial coordinate and $t$ the time. Note that the behavior (\ref{eq:wave})  may extend also beyond the horizon for negative $z-z_0$ \cite{Brout}. The dimensionless power $i\nu$ characterizes the wave singularity. The real part of $\nu$ gives the number of phase cycles per e-fold of $z-z_0$ and the imaginary part of $\nu$ describes how the amplitude grows near the horizon. For a black hole $\nu$ equals $\omega/\alpha$ where $\alpha$ denotes the gravitational acceleration at the horizon divided by the speed of light.

Logarithmic phase singularities of the type (\ref{eq:wave}) are not confined to the physics of waves near black holes. Sound waves in fluids suffer a similar fate in transsonic flows (at sonic horizons) \cite{Unruh,Visser,Garay,LKO}
and so do waves in accelerated frames (Rindler coordinates) \cite{Brout}, light in singular dielectrics \cite{Reznik} and slow light close to a zero of the group velocity \cite{Leo1,Leo2}. We refer to a behavior of the type (\ref{eq:wave}) as a {\it wave catastrophe} \cite{Leo1,Leo2}, in contrast to diffraction catastrophes \cite{BerryUpstill,Nye} that are catastrophes of light rays in the sense of catastrophe theory \cite{Thom, Arnold} as singularities of gradient maps. The ray catastrophes are accompanied by characteristic wave effects --- interference patterns, whereas the wave catastrophes may be responsible for characteristic quantum effects \cite{Leo1,Leo2,Berry1,Berry2} --- spontaneous particle production \cite{Brout,Hawking}. The spectrum of the generated quantum radiation seems to depend on the imaginary part of the index $\nu$ \cite{Leo1,Leo2}. All the examples of wave catastrophes studied so far correspond to 
\begin{equation}
\rm{Im}\, \nu = \frac{n}{2}\,, \quad n \in Z \,.
\label{eq:halfn}
\end{equation}
In this paper we develop a simple argument showing that the property (\ref{eq:halfn}) is not a coincidence for two general classes of wave catastrophes.
(A) The catastrophe affects only waves propagating in one direction, such as the outgoing waves from the horizon of a black hole where incident waves are not singular.  In this case only one non-zero $\nu$ exists at the horizon $z_0$. (B) The catastrophe affects both directions equally. This case corresponds to two powers $i\nu$ that are complex conjugate, one describing waves propagating to the right and the other refers to waves propagating to the left. Waves in accelerated frames \cite{Brout}, light in singular dielectrics \cite{Reznik} and slow-light catastrophes \cite{Leo1,Leo2} belong to case (B).

Consider real scalar waves $\phi$ in 1+1 dimensions that are subject to the Principle of Least Action. We assume that the wave equation is linear and of second order. Consequently the most general Lagrangian for $\phi$ is
\begin{equation}
{\mathscr L} = A^{\mu\mu'}
(\partial_\mu \phi) (\partial_{\mu'} \phi) +
2B^\mu \phi\, \partial_\mu \phi + C \phi^2 \,.
\end{equation}
The indices refer to the time $t$ ($\mu=0$) and to the spatial coordinate $z$ ($\mu=1$), the $\partial_\mu$ denote partial derivatives and we employ Einstein's summation convention. Without loss of generality we assume that
\begin{equation}
A^{\mu\mu'} = A^{\mu'\mu} \,.
\end{equation}
We express the Lagrangian as 
\begin{equation}
{\mathscr L} = A^{\mu\mu'}
(\partial_\mu \phi) (\partial_{\mu'} \phi) +
\phi^2 (C-\partial_\mu B^\mu) +
\partial_\mu (B^\mu \phi^2) \,.
\end{equation}
Since the divergence $\partial_\mu (B^\mu \phi^2)$ does not influence the action and hence the equations of motion, we can reduce the problem to a Lagrangian of the form 
\begin{equation}
{\mathscr L} = A^{\mu\mu'}
(\partial_\mu \phi) (\partial_{\mu'} \phi) -
M \phi^2 
\label{eq:la}\,.
\end{equation}
We obtain the Euler-Lagrange equations
\begin{equation}
\partial_\mu A^{\mu\mu'}\partial_{\mu'} \phi + M \phi = 0\,.
\end{equation}
We assume a stationary regime where $A^{\mu\mu'}$ and $M$ may depend on $z$, but not on $t$, and we decompose $\phi$ into monochromatic waves $\varphi$ with
\begin{equation}
\partial_t \varphi = -i\omega\varphi \,.
\end{equation}
We get
\begin{equation}
\Big(-\omega^2 A^{00} - 2i\omega A^{01} \partial_z - i\omega(\partial_z A^{01}) + \partial_z A^{11} \partial_z + M \Big) \varphi = 0 \,.
\label{eq:weq}
\end{equation}
In order to obtain a solution of the type (\ref{eq:wave}) near $z_0$, the $A^{\mu\mu'}$ matrix and $M$ must behave like
\begin{equation}
A^{\mu\mu'} \sim \gamma
\left(
    \begin{array}{cc}
      \frac{\displaystyle\alpha}{\displaystyle z-z_0} & \beta \\
       \beta & z_0-z 
    \end{array}
\right) (z-z_0)^n \,,\quad
M \sim \gamma\delta\, (z-z_0)^{n-1}
\label{eq:am}
\end{equation}
with the constants $\alpha$, $\beta$, $\gamma$, $\delta$. The power $n$ must be integer, because otherwise the Lagrangian (\ref{eq:la}) is not real for all $\phi$. We substitute the structure (\ref{eq:am}) into the wave equation (\ref{eq:weq}) and solve for $\nu$ with the result
\begin{equation}
\nu = i\frac{n}{2} - \beta\omega \pm \sqrt{(\alpha + \beta^2)\omega^2 - \delta^2 - \frac{n^2}{4}} \,.
\end{equation}
If the wave catastrophe affects only wave propagating in one direction, like the outgoing wave from a black hole,  one of the $\nu$ is zero. Consequently, the other is $in/2-2\beta\omega$. In the case the catastrophe affects both directions equally, $\beta$ must vanish. If $\nu$ corresponds to a wave catastrophe it must have a non-vanishing real part. Consequently, the imaginary part of $\nu$ is $n/2$. 

This proves our statement. At wave catastrophes, {\it i.e.}\ at logarithmic phase singularities, the wave amplitudes rise with the power $n/2$ for monodirectional and  for symmetric catastrophes.

Our paper was supported by 
the ESF Programme 
Cosmology in the Laboratory,
the Leverhulme Trust,
the National Science Foundation of Hungary (contract No.\ T43287),
and the Marie Curie Programme of the European Commission.


\newpage

\begin{thebibliography}{99}
\bibitem{Brout}
R. Brout, S. Massar, R. Parentani, and Ph. Spindel, Phys. Rep.
{\bf 260}, 329 (1995).
\bibitem{Unruh}
W. G. Unruh, Phys. Rev. Lett. {\bf 46}, 1351 (1981).
\bibitem{Visser}
M. Visser,
Class. Quantum Grav. {\bf 15}, 1767 (1998).
\bibitem{Garay}
L. J. Garay, J. R. Anglin, J. I. Cirac, and P. Zoller, 
Phys. Rev. Lett. {\bf 85}, 4643 (2000).
\bibitem{LKO}
U. Leonhardt, T. Kiss, and P. \"Ohberg, 
 J. Opt. B {\bf 5}, S42 (2003). 
\bibitem{Reznik}
B. Reznik, Phys. Rev. D {\bf 62}, 044044 (2000).
\bibitem{Leo1}
U. Leonhardt, 
Nature {\bf 415}, 406  (2002). 
\bibitem{Leo2}
U. Leonhardt, 
Phys. Rev. A {\bf 65}, 043818 (2002). 
\bibitem{BerryUpstill} M. V. Berry and C. Upstill, 
Prog. Optics {\bf 28} 257 (1980).
\bibitem{Nye}
J. F. Nye, {\it Natural Focusing and Fine Structure of Light}
(Institute of Physics, Bristol, 1999).
\bibitem{Thom}
R. Thom, {\it Stabilit\'{e} structurelle et morphog\'{e}n\`{e}se}
(Benjamin, Reading, 1972); English translation {\it Structural
Stability and Morphogenesis} (Benjamin, Reading, 1975).
\bibitem{Arnold}
V. I. Arnol'd, Uspekhi Mat. Nauk {\bf 30}, 3 (1975) [Russian
Math. Surveys {\bf 30}, 1 (1975)].
\bibitem{Berry1}
M. V. Berry, {\it Rays, wavefronts and phase: a picture book of
cusps}, in {\it Huygen' Principle 1690-1990: Theory and
Applications} edited by H. Blok, H. A. Frewerda, and H. K. Kuiken
(Elsevier, Amsterdam, 1992).
\bibitem{Berry2}
M. V. Berry, 
SPIE {\bf 3487}, 1 (1998).
\bibitem{Hawking}
S. M. Hawking, 
Nature {\bf 248}, 30 (1974).
\end{thebibliography}
\end{document}